\documentclass[pra,twocolumn,groupedaddress]{revtex4-1}
\usepackage{amsmath,amssymb}
\usepackage{amsthm}
\usepackage{color}

\usepackage{qcircuit}
\usepackage{graphicx}
\DeclareMathOperator{\tr}{tr}

\newcommand{\ket}[1]{\left|#1\right\rangle}
\newcommand{\bra}[1]{\left\langle#1\right|}

\newtheorem*{remark}{}
\begin{document}


\title{The SWITCH test for discriminating quantum evolutions}


\author{P. Chamorro-Posada}
\email[]{pedcha@tel.uva.es}
\author{J.C. Garcia-Escartin}
\affiliation{Dpto. de Teor\'{\i}a de la Se\~nal y Comunicaciones e Ing. Telem\'atica, Universidad de Valladolid, ETSI Telecomunicaci\'on, Paseo Bel\'en 15, 47011 Valladolid, Spain}


\date{\today}

\begin{abstract}
We propose a quantum circuit to discriminate between two arbitrary quantum evolution operators.  It permits to test the equality of two quantum operators and to estimate a fidelity measure of them.  The relation of the proposal to the SWAP test for discriminating two quantum states is analyzed.  We also discuss potential applications for the discrimination of quantum communication channels and possible laboratory implementations with light along the same lines of recent experimental realizations of quantum superpositions of causal orders exploiting the different degrees of freedom of photons. We also discuss hardware efficient realizations for noisy intermediate scale quantum computers.
\end{abstract}


\maketitle
Distinguishing between two objects is a most fundamental task in both quantum and classical information theory. For instance, discriminating two quantum channels is a key problem of quantum information \cite{acin,lopresti,max1,max2,duanseq,duanloc, memorydisc,chiribella2012,illupol,illugauss,shapiro,duanqo,harrow,montanaro2016}. In this work, we put forward a test for comparing two quantum systems.

Quantum information processing tasks are conventionally described as quantum circuits. Using quantum gates as building blocks, the evolution from an initial state is represented by a circuit. Quantum superpositions of states provide the intrinsic parallelism absent in computations performed using classical means. Nevertheless, quantum theory also permits the superposition of quantum operations \cite{aharonov1990}. One implication of this property of quantum systems is that it allows for a relaxation of the notion of a predefined causal order. Such a dynamic causal structure can be fundamental, for instance, in the physical description of quantum gravity \cite{hardy2005}. Recently, this view of a quantum theory without a definite causal structure has opened a new way to study quantum computation. The simplest example of this new form of quantum computation is a quantum switch where a qubit controls the causal order of a quantum circuit composed of two cascaded systems \cite{chiribella2013,colnaghi2012,oreshkov2016}.  

The superposition of evolutions can be simulated in a conventional quantum circuit by expanding the Hilbert space dimension with an ancillary control qubit. There exists a recent optical implementations of such a system, where the superposition of gate orders is created using additional degrees of freedom of the photon \cite{procopio2015,GGK18,RRM22}.  

The SWITCH test presented in this work also relies on this type of simulation of the superposition of quantum evolutions and can be thought of as being based on a simplified version of a quantum switch \cite{TK02,GC06b}. Even though it does not imply any modification of the causal order, it bears some resemblance with those recent proposals that explore the causal structure of quantum physics. 

These quantum switches can provide advanced communication channels with enhanced capacity and improved behaviour against noise \cite{GFA16,ESC18,GHH20,CC20,GCP20,RRE21,CBS21}.

Our suggestions for a SWITCH test is closely related to the SWAP test. Originally proposed for quantum fingerprinting applications \cite{buhrman2001}, the SWAP test permits to verify whether two quantum states are equal or not. It has also been shown to be equivalent to the Hong-Ou-Mandel effect \cite{schwarz2011,escartin2013}.  This test is based on a quantum controlled SWAP gate.  Depending on the state of the control qubit, $\left|0\right\rangle$ or $\left|1\right\rangle$, it respectively leaves unchanged the input states $\left|\phi\right\rangle\left|\psi\right\rangle$ or swaps them producing the output $\left|\psi\right\rangle\left|\phi\right\rangle$.

\begin{figure}[h]
\mbox{
\Qcircuit @C=1em @R=.7em {
& \lstick{\ket{0}} & \gate{H} & \ctrl{1} &\gate{H} & \meter \\
& \lstick{\ket{\phi}} & {/}\qw & \multigate{1}{SWAP}&\qw{/}&\qw\\
& \lstick{\ket{\psi}} & {/} \qw & \ghost{SWAP}&\qw{/}&\qw
}
}
\caption{Quantum circuit implementing the SWAP test.}\label{fig:swap}
\end{figure}
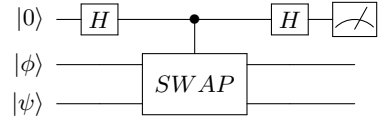

The quantum circuit implementing the SWAP test is shown in Fig. \ref{fig:swap}.  The aim is to test whether quantum states $\left|\psi\right\rangle$ and $\left|\phi\right\rangle$  are equal or not.  Before the measurment of the ancillary qubit, the resulting quantum state is given by
\begin{equation}
\frac{1}{2}\left[\left|0\right\rangle\left(\left|\phi\right\rangle\left|\psi\right\rangle+\left|\psi\right\rangle\left|\phi\right\rangle\right)+\left|1\right\rangle\left(\left|\phi\right\rangle\left|\psi\right\rangle-\left|\psi\right\rangle\left|\phi\right\rangle\right)\right]. \label{eq:swap}
\end{equation}

The SWAP test is passed if the measurement of the ancillary qubit gives $0$ and fails otherwise. If the two states are equal, $\left|\psi\right\rangle=\left|\phi\right\rangle$, the test is always passed.  When the states are different, there is a finite probability $P$ of still passing the test that depends on the overlap of the two states $\left|\left\langle\psi|\phi\right\rangle\right|^2$
\begin{equation}
P=\frac{1}{2}\left(1+\left|\left\langle\psi|\phi\right\rangle\right|^2\right).\label{eq:Pswap}
\end{equation}
Even for very similar states with a large overlap, the probability of two states that are different passing the test becomes exponentially small in the number of repetitions of the test and the equality can be verified with any required confidence if a sufficiently large number of copies of the states are available. These results can be easily extended to mixed states. For two mixed states with density matrices $\rho$ and $\sigma$, the probability of passing the SWAP test is $\frac{1+\tr{(\rho\sigma)}}{2}$ \cite{kobayashi2003}.

\begin{figure}[h]
\mbox{
\Qcircuit @C=1em @R=.7em {
& \lstick{\ket{0}} & \gate{H} & \ctrl{1} &\gate{H} & \meter \\
& \lstick{\ket{\phi}} & \qw{/ } & \gate{SWITCH\lbrace U_1,U_2\rbrace}&\qw{/}&\qw\\
}
}
\caption{Quantum circuit implementing the SWITCH test.}\label{fig:switch}
\end{figure}
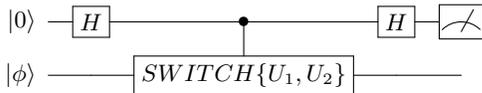

We present similar tests for quantum operators. Fig. \ref{fig:switch} shows a basic SWITCH test that helps to understand the possibilities and limitations of our approach. This is a modified version of the Hadamard test used in quantum algorithms for the calculation of Jones polynomials \cite{AJL06,PMR09} or in different quantum machine learning tasks \cite{BLS19,CWV20}. For an arbitrary input state $\left|\phi\right\rangle$ and $U_1$ and $U_2$ the unitary evolutions being tested, we define the output of the controlled $\text{SWITCH}\left\lbrace U_1,U_2\right\rbrace$ gate as $U_1\left|\phi\right\rangle$ if the control qubit is $\left| 0 \right\rangle$ and $U_2\left|\phi\right\rangle$ if it is $\left| 1 \right\rangle$.  This permits to create the superposition of the two evolutions $U_1$ and $U_2$. 

Before the measurement of the ancillary qubit, the resulting quantum state is given by   
\begin{equation}
\ket{\varphi}=\frac{1}{2}\left[\left|0\right\rangle\left(U_1\left|\phi\right\rangle+U_2\left|\phi\right\rangle\right)+\left|1\right\rangle\left(U_1\left|\phi\right\rangle-U_2\left|\phi\right\rangle\right)\right]. \label{eq:switch}
\end{equation}

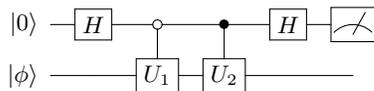
\begin{figure}[h]
\mbox{
\Qcircuit @C=1em @R=.7em {
& \lstick{\ket{0}} & \gate{H} & \ctrlo{1} & \ctrl{1} &\gate{H} & \meter \\
& \lstick{\ket{\phi}} & {/}\qw & \gate{U_1}&\gate{U_2}&\qw{/}&\qw\\
}
}
\caption{Quantum circuit implementation of a SWITCH gate test with one test state.}\label{fig:switch2}
\end{figure}

\begin{figure}[h]
\includegraphics[width=0.8\columnwidth]{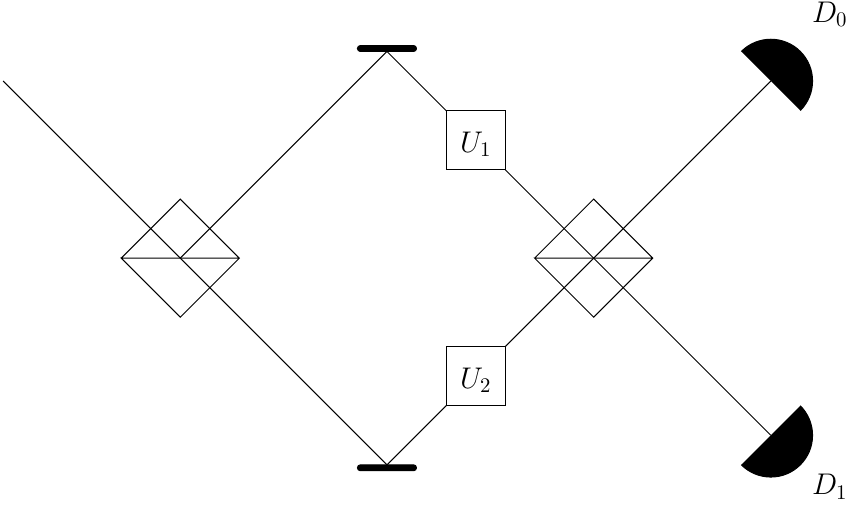}
\caption{Optical implementation of a single state SWITCH test with an interferometer. The test is passed if detector $D_0$ measures a photon.}\label{fig:optswitch}
\end{figure}

A circuit diagram of a possible implementation, when the two operators $U_1$ and $U_2$ can be made subject to a quantum control, is shown in  Fig. \ref{fig:switch2}. A simple deployment of this system for an optical system could follow the same lines demonstrated in \cite{procopio2015} with the control qubit realized spatially and the quantum operations acting on the polarization degree of freedom of the same photon. The system can be implemented using a Mach-Zehnder interferometer as shown in Figure \ref{fig:optswitch}. A photon is sent through a $50/50$ beam splitter that creates the spatial qubit implementing the first Hadamard gate in the control channel in Fig. \ref{fig:switch2}. State  $\left|0\right\rangle$ is obtained if it is reflected at the beamsplitter and $\left|1\right\rangle$ if it is transmitted.  The two spatial paths are associated with two different operations on the photon polarization $U_1$ and $U_2$. At the output, the two paths recombine coherently at a second $50/50$ beamsplitter that implements the second Hadamard gate in Fig. \ref{fig:switch2}. The measurement of the photon and the determination of the output path completes the SWITCH test for the quantum operations $U_1$ and $U_2$. 
This kind of interferometric comparison setup was already proposed in \cite{andersson2003} as a way to tell apart two unknown evolutions.

This system can compare qubits encoded in the polarization degree of freedom, but we can have access to higher-dimensional systems if we take advantage of other degrees of freedom of the photon, like its orbital angular momentum. With these encodings, each photon will represent an input state in a larger Hilbert space.

The SWITCH test is passed if the measurement of the ancillary qubit gives $0$ and fails otherwise. If the two systems are equal, $U_1=U_2$, the measurement result  is necessarily  $0$ and the test is always passed.  When the systems are different, there is a finite probability $P$ of still passing the test that is dependent on the input state $\left|\phi\right\rangle$. The probability of passing the test can be written in terms of the trace of the product of the measurement projector and the density matrix representation of the state $\ket{\varphi}$ in Eq. (\ref{eq:switch}), $\tr(\ket{0}\bra{0}\ket{\varphi}\bra{\varphi})$, which gives 
\begin{equation}
P=\frac{1}{2}\left(1+ \text{Re}\{\tr{(U_2\rho U_1^\dag)}\}\right)\label{eq:probTr}
\end{equation}
or
\begin{equation}
P=\frac{1}{2}\left(1+\text{Re}\left\lbrace \left\langle\phi\right|U_1^\dag U_2\left|\phi\right\rangle\right\rbrace\right) \label{eq:prob}
\end{equation}
for pure states.

A failure of the test, under ideal conditions, confirms that the two systems $U_1$ and $U_2$ are certainly distinct. Passing the test gives some probability that the systems are identical. This probability can be optimized by the repetition of the test a sufficient number of times, but it requires a careful strategy.    

There are some clear limitations in the SWITCH test as proposed so far. For instance, it can happen that $U_1\left|\phi\right\rangle=U_2\left|\phi\right\rangle$ and a repeated use of the state $\left|\phi\right\rangle$ would always pass the test, even though the operators are different. A clear example for two qubit states are the CNOT operation, $CNOT\ket{x}\ket{y}=\ket{x}\ket{x+y \mod 2}$, and the identity ($U_1=CNOT, U_2=I$). Both unitaries would leave a state $\ket{\phi}=\ket{00}$ unchanged. Therefore, as opposed to the SWAP test, a repeated evaluation of the SWITCH test with the same input state is not a judicious option. 

We need an appropriate collection of states $\left\lbrace \left|\phi_i\right\rangle\right\rbrace$ to be used in multiple test repetitions. For instance, we could use a basis spanning the Hilbert space on which the $U_j$ gates act. If the operators are different, there will be at least one input state of the basis for which the output states are different and there is a probability greater than 0 of failing the test. This is too costly.

A good alternative is sampling random input states. If we choose uniformly at random a unitary $U_r$ (sampling from the Haar measure), we can apply it on a fixed initial state, for instance setting all the qubits to $\ket{0}$. The result is a random initial state. 

For a large enough repetitions of the SWITCH test, different unitaries $U_1$ and $U_2$ will eventually fail the test. We can show (see Section A.1 in the Appendix) that the average probability of passing the test with random input states becomes 
\begin{equation}
E[P]_{\text{Haar}}=\frac{1+\text{Re}\{\langle U_1,U_2 \rangle\}/d}{2}
\end{equation}
where $\langle U_1,U_2 \rangle=\tr(U_1^\dag U_2)$ is the Frobenius inner product of the two evolutions. This average gives a good estimation of how close the two operators are. 

For $U_1=U_2$, the test is always passed. For different evolutions, the average probability of passing the test will, in general, be close to 1/2 for large Hilbert spaces unless the operators are particularly close to each other. 

The random Haar unitaries we need can be generated from classical random distributions \cite{Mez07} and then converted into quantum circuits, but that can require a large number of elementary quantum gates. In the Noisy Intermediate Scale Quantum Computers that are available today \cite{Pre18}, we prefer quantum circuits of a small depth, i.e. the total number of consecutive elementary gates should be as small as possible. In longer circuits the noise at each gate can compound to the point of making the resulting output states unusable. 

There are economical approximations to random Haar matrices. For all the functions we will need to average in this paper, we can use 2-designs to obtain the same averages as the full Haar measure \cite{AE07,GAE07} using only a finite set of gates to generate the random states. We can further simplify the testing by using approximate designs \cite{HL09,DCE09}, which are enough for our purposes. 

The number of states we need to generate to obtain an exact average can grow exponentially with the size of the state space. Usually, we will take a Monte Carlo approach assuming that, for a large enough number of properly sampled random states, the finite average approximates the Haar average for a large enough number of samples. 

There are explicit compact contructions to produce pseudorandom quantum states with few gates \cite{EWS03} which can give an efficient Monte Carlo SWITCH test for our evolutions $U_1$ and $U_2$. We can also try any of the parametrized circuits used in quantum machine learning to explore the state space. These circuits generate a trial state, or \emph{ansatz} $\ket{\psi}_{\vec{\theta}}$ that depends on a short list of classical parameters $\vec{\theta}=(\theta_1,\ldots,\theta_k)$ \cite{BLS19}. A good \emph{ansatz} circuit needs to balance the ability to cover as much as the Hilbert space as possible with low number of stages so that the noise cannot build up to a dangerous level. For the proposed use for pseudorandom state generation, reasonable \emph{ans\"atze} should be good options to provide random sampling. 

These sampling strategies can be taken to the interferometric setups we have discussed for optical channels. There are efficient optical systems that generate random Haar evolutions \cite{RCO17} as well as configurable integrated optical circuits that can be adjusted on demand \cite{PMS14,EPS20,TMS21,HPG22}.

Apart from the input state problem, the test is dependent on a phase reference. For instance, testing $U$ against $e^{i\alpha}U$ for $\alpha=\frac{\pi}{2}$ would always give a negative result even though the two operators would produce states that are equal up to a global phase and, as such, indistinguishable. However, using controlled gates introduces a phase reference. The control qubit allows to distinguish relative phases that would be unmeasurable if we only had the systems under test. This is, in fact, the expected behaviour for interferometers, which are used to find phase differences that serve as indirect measurements of other magnitudes, such as changes in length. Our system is equivalent to an interferometer and must show a phase dependence, which is associated to the real part in equation \eqref{eq:prob}. 

The dependence on the phase can be somewhat countered if the test is repeated for $U_1$, $U_2$ and a phase-shifted version of $U_1$, $iU_1$, and $U_2$. The second test permits to estimate $\text{Im} \lbrace \left\langle\phi\right|U_1^\dag U_2\left|\phi\right\rangle\rbrace$. If we combine both tests, we can reconstruct an estimate of $|  \left\langle\phi\right|U_1^\dag U_2\left|\phi\right\rangle|^2$. Of course, we would need to repeat the measurement for a suitable set of test states $\left\lbrace\left|u_i\right\rangle\right\rbrace$ and we would loose the desirable property that a negative tells us for sure the systems are different when there appear phase shifts.

\begin{figure}[h]
\mbox{
\Qcircuit @C=1em @R=.7em {
& \lstick{\ket{0}} &  \qw& \gate{H} & \ctrl{1} &\gate{H} & \meter \\
& \lstick{\ket{\phi}} &{/} \qw&\gate{U_1} & \multigate{1}{SWAP}&\qw{/}&\qw\\
& \lstick{\ket{\psi}} &{/} \qw&\gate{U_2} & \ghost{SWAP}&\qw{/}&\qw
}
}
\caption{Quantum circuit implementing the modified SWAP test.}\label{fig:swap2}
\end{figure}
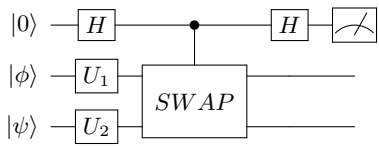

In order to remedy these shortcomings, we can modify the SWAP test as shown in Fig. \ref{fig:swap2}. In the repeated operation of the circuit, the two inputs are set to equal values. For an input density matrix $\rho$ that is a linear combination of matrices of the form $\ket{\phi}\ket{\phi}$, the SWAP test succeeds with a probability
\begin{equation}
P=\frac{1+\tr{(U_2\rho U_2^\dag U_1\rho U_1^\dag)}}{2}\label{eq:Pswapswitch}
\end{equation}
or
\begin{equation}
P=\frac{1}{2}\left(1+\left|\bra{\phi}U_1^\dag U_2\ket{\phi}\right|^2\right).\label{eq:Pswapswitchpure}
\end{equation}
for pure state inputs of the form $\ket{\phi}\ket{\phi}$.

We can compute again the average probability of passing this test for random Haar input states where we generate two copies of the same state for each of the two inputs of the circuit. The result (see Section A.2 in the Appendix) is:
\begin{equation}
E[P]_{\text{Haar}}= \frac{d+|\tr(U_1^\dag U_2)|^2}{d(d+1)}.
\end{equation}
This probability is proportional to the squared absolute value of the Frobenius inner product of the operators and gives an idea of how one unitary ``projects'' into the other. 
As required, when $U_1=U_2$, the test is passed for any input state.

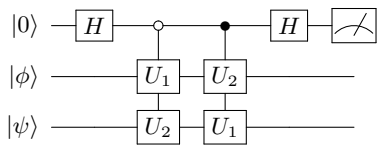
\begin{figure}[h]
\mbox{
\Qcircuit @C=1em @R=.7em {
& \lstick{\ket{0}} & \gate{H} & \ctrlo{1} & \ctrl{1} &\gate{H} & \meter \\
& \lstick{\ket{\phi}} & \qw{/} & \gate{U_1}&\gate{U_2}&\qw{/}&\qw\\
& \lstick{\ket{\psi}} &\qw{/}& \gate{U_2} \qwx &\gate{U_1} \qwx&\qw{/}&\qw\\
}
}
\caption{Quantum circuit implementation of a two-state SWITCH gate test.}\label{fig:switch3}
\end{figure}

For the purpose of comparing operators, the alternative SWITCH test circuit of Fig. \ref{fig:switch3} is equivalent to the modified SWAP test.  Even though the general evolution is different from that of the circuit in Fig. \ref{fig:swap2}, the outputs are identical when input states $\left|\phi\right\rangle=\left|\psi\right\rangle$ are used.

\begin{figure}[h]
\includegraphics{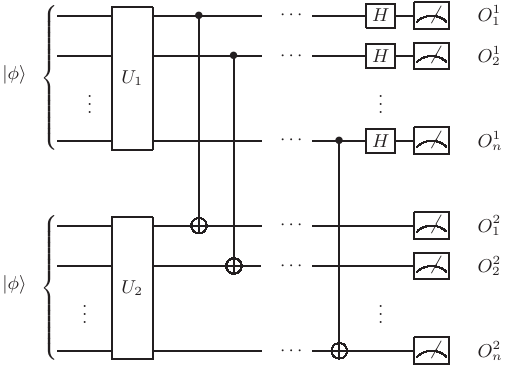}
\caption{Quantum circuit implementation of a destructive two-state SWITCH test. The circuit complexity besides the unitaries $U_1$ and $U_2$ can be reduced to a measurement circuit with two stages and light classical processing.}\label{destructiveSWAP}
\end{figure}

The cost of this approach is doubling all the qubits except for the control qubit. However, this modification allows for a reduction in the number of gates in the test. Usually a controlled unitary requires multiple two qubit gates and there are different decomposition strategies. This is a general problem in the Hadarmard test of Figure \ref{fig:switch} and derived measurements and there have been various attempts to reduce the depth of these circuits \cite{LLS08,MF19,AA21,WRZ21}. In the SWAP test this problem can be circumvented. It is possible to perform a SWAP test without a controlled SWAP. The circuit in Figure \ref{destructiveSWAP} shows the destructive SWAP test presented in \cite{GC13} and later rediscovered by a machine learning algorithm \cite{CSS18}. The additional cost of this test is performing some simple classical computations on the measurement results. The test reduces to performing a Bell measurement on each corresponding pair of qubits of the states to be compared and then computing the parity of the bitwise AND of the obtained binary sequences (assigning the 0 bit to a $\ket{0}$ result and 1 to $\ket{1}$). The whole cost, apart from the unavoidable gates for $U_1$ and $U_2$, is adding $n$ Hadamard gates and $n$ CNOT gates. Classically, we just need simple AND operations and counting the number of 1s in the resulting binary sequence (a series of modulo 2 additions). In noisy implementations, this depth reduction can compensate for the additional number of qubits. 

\begin{figure}[h]
\includegraphics[width=0.4\columnwidth]{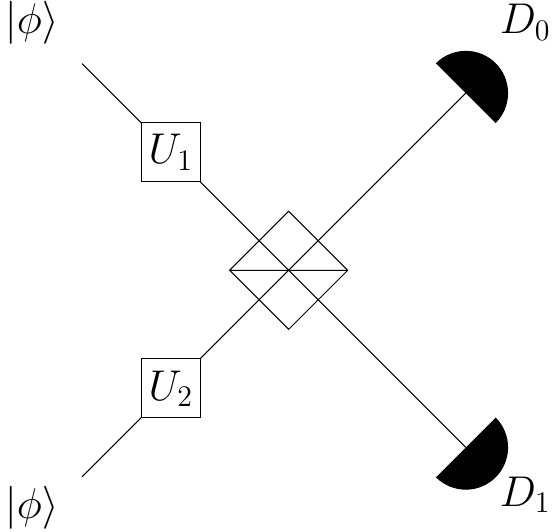}
\caption{Optical implementation of a two-state SWITCH test with Hong-Ou-Mandel interference. The test fails if we find a coincidence (both detectors measure a photon at the same time).}\label{fig:optswitchHOM}
\end{figure}

Finally, we would like to show some possible optical implementations of these two-state SWITCH tests. We assume the probe states are qubits encoded into single photons using their different degrees of freedom such as polarization, frequency or orbital angular momentum. Figure \ref{fig:optswitchHOM} shows an optical setup implementing the circuit of Figure \ref{fig:swap2}.

The comparison comes from the quantum interference of single photons at a balanced beamsplitter in the Hong-Ou-Mandel experiment \cite{hong87} which, for input states with density matrices $\rho$ and $\sigma$, gives a probability $P=\frac{1-\tr{(\rho \sigma)}}{2}$ of finding a coincidence. Two photons in the same state will bunch and always come out together, which gives a natural way to implement a SWAP test \cite{GC13}. If we can produce any state on demand and have two operations we want to compare, the presented optical setup implements a SWITCH test. 

We have shown the proposed SWITCH tests can estimate how close two quantum evolution operators are. The one-state test has several limitations, but can be readily realized in interferometric experiments with one photon. The two-state tests are related to the SWAP test used to compare quantum states. We have given two alternative quantum circuits implementing this test, one of which can be implemented in a quantum optics lab using a beamsplitter in a Hong-Ou-Mandel setup. This expands previous optical setups that can compare quantum processes using entanglement \cite{LRO09}. We have also commented different sampling strategies to obtain meaningful averages. The required random state generation can be performed with simple quantum circuits which could be built with the current noisy intermediate scale quantum computers. In optical setups, there are also known constructions that allow to prepare the needed input states. Combined with the recent advances in integrated photonics, this offers a compact way to compare to two optical quantum communication channels. 

\begin{acknowledgments}
The authors acknowledge support from the Spanish Ministerio de Ciencia e Innovaci\'on (MCIN), project PID2020-119418GB-I00, and funding from the European Union NextGenerationEU (PRTRC17.I1) and the Consejer\'ia de Educaci\'on, Junta de Castilla y Le\'on, through QCAYLE project.
\end{acknowledgments}

\renewcommand{\thesection}{\Alph{section}}
\renewcommand{\thesubsection}{\Alph{section}.\arabic{subsection}}
\section{Appendix}
In our tests, we want to consider the average of functions $f(V): V \rightarrow \mathbb{C}$ acting on unitary matrices $V$ chosen at random with respect to the Haar measure. This average is given by the integral
\begin{equation}
E\left[f(V)\right]_\text{Haar}=\int_{V\in \mu_{V}} f(V) d\mu_{V},
\end{equation}
where $\mu_{V}$ is the Haar measure for the unitary group we are interested in. 

There are two common computation tricks that will be useful. In both cases we consider two Hilbert spaces of dimension $d$, $\mathcal{H}^d_1$ and $\mathcal{H}^d_2$, and the joint product space $\mathcal{H}=\mathcal{H}^d_1\times \mathcal{H}^d_2$. 

First, we notice that, for the trace operation and two unitary matrices $A,B$ acting on states in $\mathcal{H}^d$,
\begin{equation}
\tr(AB)=\tr(\text{SWAP}(A\otimes B)),
\label{SWAPtrick}
\end{equation}
where the SWAP operation acts on the halves corresponding to each individual space, $\mathcal{H}^d_1$ and $\mathcal{H}^d_2$, and $\otimes$ is the usual tensor product. 

This SWAP trick \cite{DPS12} can be easily verified by writing the tensor product as a sum of the weights for each matrix position so that 
\begin{equation}
\begin{split}
&\tr(\text{SWAP}(A\otimes B))=\tr\!\left(\!\text{SWAP}\!\sum_{ijkl} A_{ij} B_{kl}\ket{i}\bra{j}\otimes \ket{k}\bra{l}  \right) \\
&=\sum_{ijkl} A_{ij} B_{kl}\tr\left(\ket{k}\bra{j}\otimes \ket{i}\bra{l} \right)= \sum_{ijkl} A_{ij} B_{kl}\delta_{kj}\delta_{il}\\
&=\sum_{ik} A_{ik} B_{ki}=\tr(AB).
\end{split}
\end{equation}

We will also consider the Haar average for the tensor product of a random unitary matrix and its inverse (see Corollary 3.5 in \cite{Zha14}):
\begin{equation}
\int_{V\in \mu_{V}} V\otimes V^\dag d\mu_{V}=\frac{\text{SWAP}}{d},
\label{integraltrick}
\end{equation}
where SWAP is the $2d\times 2d$ SWAP operation acting on each half of larger space. 

\vspace{1ex}
\subsection{$E[\tr(U_1^\dag V^\dag\rho_0 V U_2 )]_{\text{\normalfont Haar}}$}
\label{Secstateaverage}
We will compare the unitary matrices $U_1$ and $U_2$ using random initial states generated from the fixed state $\ket{0}$ with density matrix $\rho_0=\ket{0}\bra{0}$ under a random unitary $V$. 

First, we use the SWAP trick of Eq. (\ref{SWAPtrick}) to notice that
\begin{equation}
\begin{split}
&\tr(U_2 V \rho_0 V^\dag U_1^\dag)=\tr(V\rho_0 V^\dag U_1^\dag U_2)\\
&=\tr(\text{SWAP} ((V\otimes V^\dag) ( \rho_0 \otimes U_1^\dag U_2)).
\end{split}
\end{equation}

The average for random input states can be calculated using Eq. (\ref{integraltrick}). We find that:
\begin{equation}
\begin{split}
&E\left[\tr(U_2 V \rho_0 V^\dag U_1^\dag) \right]_\text{Haar}=\\
&= \tr\left(\text{SWAP} \int_{V\in \mu_{V}} (V\otimes V^\dag) ( \rho_0 \otimes U_1^\dag U_2) d\mu_{V}\right) \\
&= \tr\left(\text{SWAP}\frac{\text{SWAP}}{d}( \rho_0 \otimes U_1^\dag U_2)\right)\\
&=\frac{1}{d}\tr(\rho_0)\tr(U_1^\dag U_2)=\frac{\tr(U_1^\dag U_2)}{d},
\end{split}
\label{overlapAv}
\end{equation}
for our pure state input $\rho_0$ of trace 1.

\subsection{$E[\bra{0}V^\dag U_1^\dag U_2 V\ket{0}\bra{0}V^\dag U_2^\dag U_1 V\ket{0}]_{\text{\normalfont Haar}}$}
\label{SecHaarsquare}
We will compute the Haar average of the norm $|\bra{\psi}U_1^\dag U_2\ket{\psi}|^2$ using the results from reference \cite{Zha14}.
\begin{remark}[Corollary (3.13) of \cite{Zha14}:]
For two endomorphisms $X, Y \in \text{End}(\mathbb{C}^d)$ taking quantum states into quantum states, the average of $\bra{\psi}X\ket{\psi}\bra{\psi}Y\ket{\psi}$ over random state vectors $\ket{\psi}$ on the unit sphere $\mathbb{S}^{2d-1}$ in $\mathbb{C}^d$ is given by:
\begin{equation}
\int_{\mathbb{S}^{2d-1}}\bra{\psi}X\ket{\psi}\bra{\psi}Y\ket{\psi}d\ket{\psi}=\frac{\tr(XY)+\tr{X}\tr{Y}}{d(d+1)}.
\label{stateaverage}
\end{equation}
\end{remark}

For unitaries $U_1$ and $U_2$:
\begin{equation}
\begin{split}
&E[\bra{\psi}U_1^\dag U_2\ket{\psi}\bra{\psi}U_2^\dag U_1\ket{\psi}]_{\text{Haar}}\\
&=\frac{\tr(U_1^\dag U_2 U_2^\dag U_1)+\tr(U_1^\dag U_2)\tr(U_2^\dag U_1)}{d(d+1)}\\
&=\frac{d+|\tr(U_1^\dag U_2)|^2}{d(d+1)}.
\end{split}
\label{HaarSquare}
\end{equation}
For $U_1=U_2$, $|\tr(U_1^\dag U_2)|^2=d^2$ and the test succeeds with probability one.

While we have considered unitaries, this result uses Eq. (\ref{stateaverage}), which is valid for completely general quantum transformations, showing the SWITCH test can be used to compare any arbitrary pair of quantum channels. 

The result from Eq. (\ref{HaarSquare}) can also be deduced using Proposition 3.9 of \cite{Zha14} and using the SWAP and the integration tricks in a similar fashion to what was done in Eq. (\ref{overlapAv}).


\newcommand{\noopsort}[1]{} \newcommand{\printfirst}[2]{#1}
  \newcommand{\singleletter}[1]{#1} \newcommand{\switchargs}[2]{#2#1}
\end{document}